\documentclass{optica-article}

\journal{opticajournal} 

\articletype{Research Article}

\usepackage{lineno}

\begin{document}

\title{Ultralow-Noise Optical Parametric Amplifier for Stimulated Raman Scattering Imaging}

\author{Ines Martin\authormark{1,2}, Yoann Pertot\authormark{1}, Olivier Albert\authormark{1}, Thomas Pinoteau\authormark{1}, Sébastien Coudreau\authormark{1}, Simon Pomarède\authormark{1}, José Villanueva\authormark{1},  Grégory Gitzinger\authormark{1}, Penda Leye\authormark{1}, Ivan Jelicic\authormark{1}, Cyril Vaneph\authormark{1}, Daniel Kaplan\authormark{1}, Simone Bux\authormark{1}, Nicolas Forget\authormark{1}, Thibaut Sylvestre\authormark{3}, Samuel Métais\authormark{2}, and Hervé Rigneault\authormark{2}
}

\address{
\authormark{1}Fastlite, 165 route des Cistes, 06600 Antibes, France\\
\authormark{2}Aix Marseille Univ, CNRS, Centrale Marseille, Institut Fresnel, Marseille, France\\
\authormark{3}Institut FEMTO-ST, UMR 6174 CNRS-Université de Franche-Comté, 25030 Besançon, France
}

\email{\authormark{*}forget@fastlite.com} 


\begin{abstract*} 
We present a 40-MHz ultrafast optical parametric amplifier (OPA), tunable from 0.8 to 1\,µm, with a relative intensity noise (RIN) matching the shot-noise floor (-160\,dB/Hz) above 2\,MHz. The OPA is pumped by a 20-W Kerr-lens mode-locked Ytterbium laser and seeded by a a supercontinuum generated in an all-normal-dispersion (ANDi) fiber. With an average output power >1.5\,W, this compact and simple scheme is an attractive alternative to synchronously-pumped optical parametric oscillators, especially within the context of stimulated Raman scattering (SRS) imaging. To illustrate the latter, we perform chemical imaging of vinegar droplets in oil by SRS microscopy.
\end{abstract*}

\section{Introduction}
Stimulated Raman scattering (SRS) microscopy is a powerful and fast-evolving vibrational imaging technique that can map the distribution of chemical bonds in three-dimensional space and real time \cite{cheng2015}. SRS bio-imaging was first achieved by using a high-repetition-rate picosecond laser source and a high-frequency modulation transfer detection scheme to achieve high sensitivity and speed, despite the low photo-damage threshold of biological systems \cite{freudiger2008}.
SRS typically uses two synchronized pulsed lasers, one known as the pump beam and the other as the Stokes beam, to coherently excite the selected molecular vibration. In SRS microscopy, both laser beams are spatially and temporally overlapped, and focused collinearly into a diffraction limit spot inside the sample. By virtue of energy conservation, vibrational excitation is accompanied by optical loss in the pump beam (stimulated Raman loss, SRL) and optical gain in the Stokes beam (stimulated Raman gain, SRG),
respectively.
Hence, SRS signal is detected as a relative intensity change of the incident laser ($\Delta I/I \simeq 10^{-3}-10^{-7}$ \cite{rigneault2018tutorial}). To achieve sensitive detection above the low-frequency laser background noise, a high-frequency (multi-MHz) modulation is applied to one laser beam and the modulation transfer to the other beam is measured to extract SRS signal using a radio-frequency (RF) lock-in amplifier. This removes the noise from the slow laser intensity fluctuation and may achieve near shot-noise-limited sensitivity if the RIN of the pump beam is shot-noise-limited at the modulation frequency - otherwise one has to resign to balanced detection with a 3\,dB penalty on sensitivity \cite{ozeki2009analysis}.
With the SRS signal being registered at each pixel, three-dimensional images can be readily acquired by raster-scanning the laser focus across the samples \cite{freudiger2014stimulated} using a laser-scanning microscope in either transmitted or reflected light directions.
SRS is also compatible with confocal and two-photon excited fluorescence microscopy, as well as other nonlinear imaging techniques, such as second harmonic generation, to achieve multi-modal optical imaging. However, full-scale multi-modal imaging implictly requires some degree of wavelength tunability, of either the pump or Stokes beam, to be able to target multiple vibrational bands and/or a variety of chromophores \cite{you2018intravital}. Although SRS excitation using two narrowband picosecond (ps) lasers has been commonly used for imaging samples with known components and well-resolved Raman frequencies, the full chemical information cannot be retrieved without scanning through a broad range of Raman frequencies. Among the SRS sequential wavelength tuning schemes, the spectral-focusing \cite{Gershgoren:03,Hellerer:04, Audier:20} method is gaining popularity by linearly chirping two broadband laser pulses and then scanning the time delay. Working with broadband optical pulses for both the pump and Stokes beams is all the more relevant as low-Raman imaging may be performed with minimal changes in the same optical setup \cite{Raanan:2019b}.

An ideal driving source for SRS and multi-modal imaging should thus gather the following properties: synchronized broadband pulses, tunable Stokes or pump wavelength, a repetition rate in the range of 40-80\,MHz, shot-noise-limited RIN at the modulation frequency, Watt-level average power. Up to now, the sources of choice for SRS have been either synchronously-pumped optical parametric oscillators (SPOPOs) \cite{brustlein2011optical} or a tandem of synchronized ultrafast lasers (Ti:Saphir and Yb lasers for example) \cite{potma2002high}. If SPOPOs driven by mode-locked Yb-fiber lasers have been established as reliable sources of high-repetition-rate picosecond pulses \cite{zad14}, ultrafast SPOPOs require long-term synchronization between the OPO cavity length and the input pump pulse repetition rate, which also dictates the minimum system size. Moreover, as a resonant device, SPOPOs require a cavity comprising specially coated mirrors to provide optical feedback at the signal and/or idler wavelengths, and of suitable focal length to enable optimum spatial mode-matching between the resonant parametric wave(s) and the input pump beam, which result in relatively high overall system complexity and cost. For many applications, SPOPOs also require active cavity length control for maximum output power and spectral stability \cite{kum14}, further increasing operating demands. More recently, fiber opo (FOPO) have reached a maturity that has enabled coherent Raman (CARS and SRS) imaging with close to shot-noise-limited RIN. However the output power remains limited \cite{wurthwein2021multi}.

 Optical parametric amplifiers (OPAs) seeded with a coherent supercontinuum (SC) are powerful tools for multi-modal nonlinear applications. Thanks to single-pass amplification, their optical scheme is not only simpler and more compact but is much also less constrained by the repetition rate of the pump laser and immune to frequency drifts. The Achilles' heel of SC-seeded OPAs at tens of MHz is the threefold: (1) a minimal pulse energy is necessary to trigger SC generation, (2) a near octave-spanning spectral bandwidth is required to take advantage of the full amplification bandwidth of the OPA, (3) the pulse-to-pulse stability and coherence of the SC is critical, especially for SRS. Even if the invention of photonic crystal fibers has greatly enhanced the efficiency of supercontinuum generation \cite{dud06}, most commercial SC sources, which are based on noise-seeded soliton dynamics, are inherently highly unstable, often unpolarized and do not exhibit the required temporal and/or polarization coherence. The key to this issue of generating fully coherent low-noise SC is to use all-normal dispersion (ANDi) polarization-maintaining (PM) fibers, which enable pulse-preserving and flat-top coherent SC generation with excellent low noise and polarization properties \cite{gen19,gen21}.

In this work, we present a supercontinuum-seeded OPA, pumped by the second harmonic of a Kerr-lens mode-locked Ytterbium oscillator at 40\,MHz. The OPA delivers a W-level signal beam, tunable from 0.8\,µm to 1\,µm, with a shot-noise-limited RIN above 2\,MHz. As a proof-of-concept, we apply SRS microscopy to the chemical imaging of droplets of water in olive oil.

\section{Tunable OPA}

\begin{figure}[h]
\centering
\includegraphics[width=0.7\textwidth]{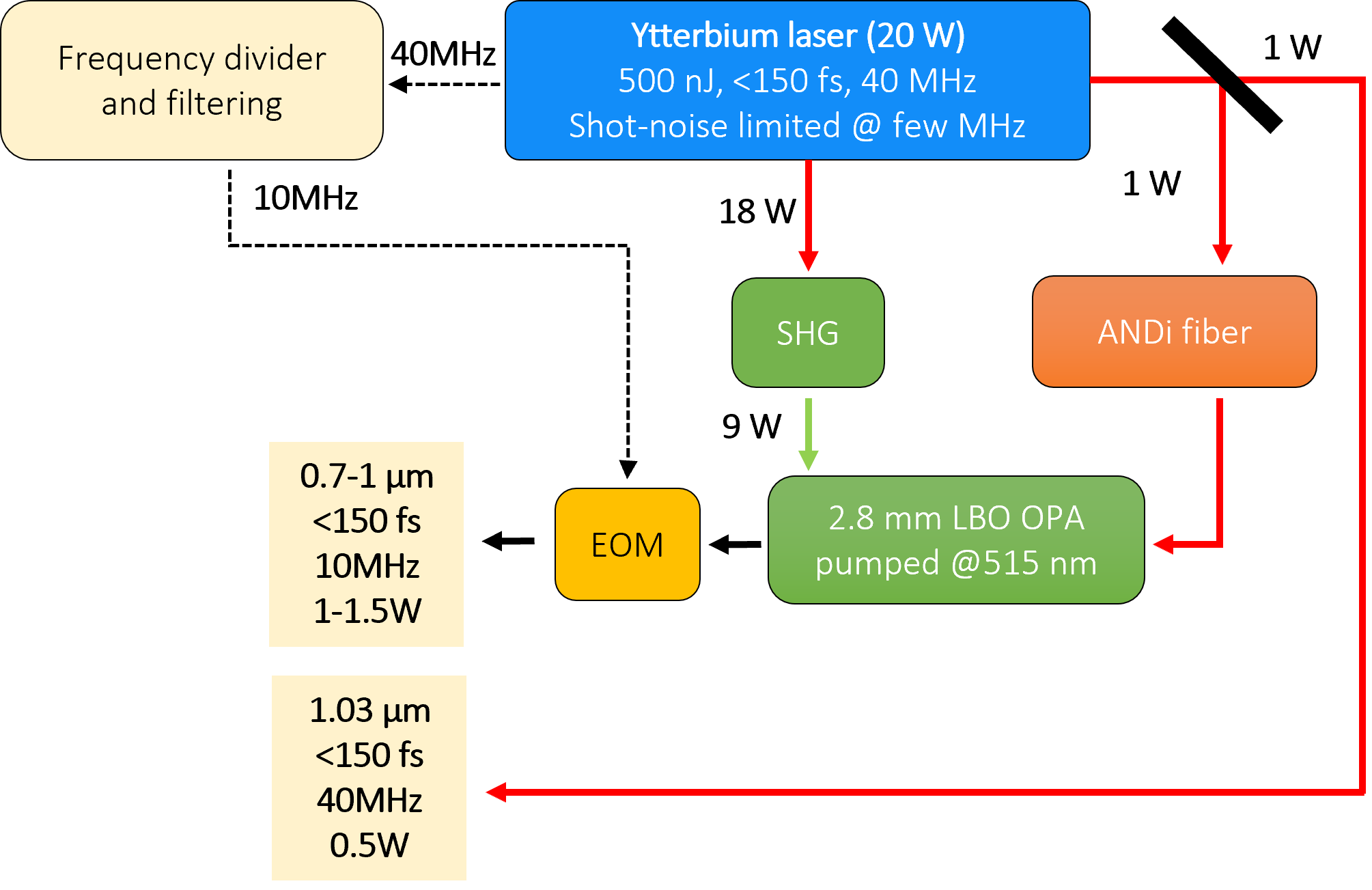}
\caption{\label{fig:OPA_setup} OPA setup }
\end{figure}

The experimental setup is shown in \autoref{fig:OPA_setup}. We employ a high-average-power Kerr-lens mode-locked Ytterbium oscillator (Flint, Light Conversion) delivering 165\,fs pulses at 1030\,nm with a repetition rate of 40\,MHz. The output average power of 20\,W corresponds to a pulse energy of 500\,nJ. Approximately 1\,W of the laser power is sampled to generate a supercontinuum in an 10\,cm long PM-ANDi photonic fiber \cite{gen21} while 18\,W is used to pump the OPA. The remaining pump power (1\,W) is saved for SRS. 

\begin{figure}[h]
\centering
\includegraphics[width=0.7\textwidth]{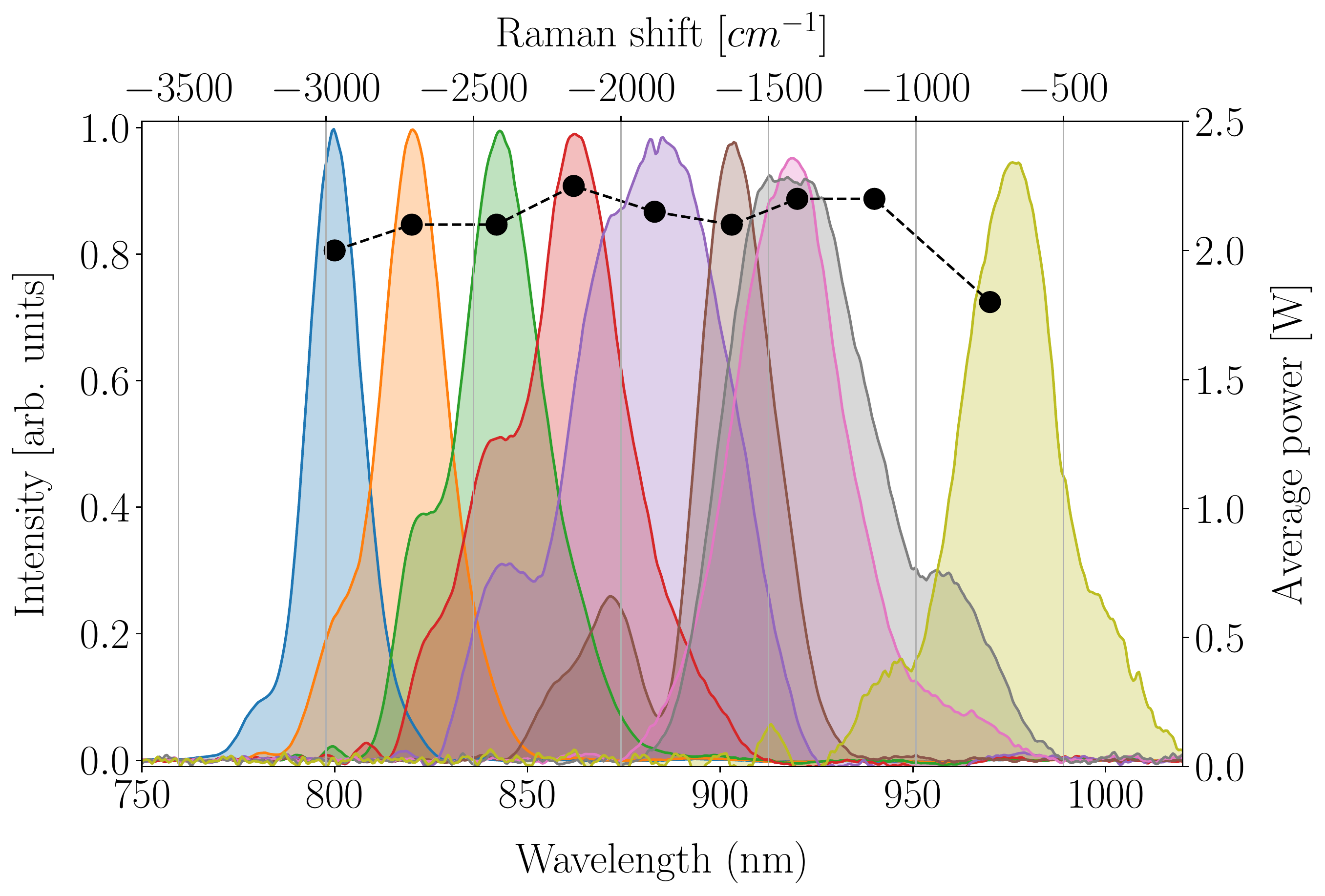}
\caption{\label{fig:OPA_spectrum} Output spectrum and output average power from the OPA.}
\end{figure}

\begin{figure}[h]
\centering
\includegraphics[width=0.7\textwidth]{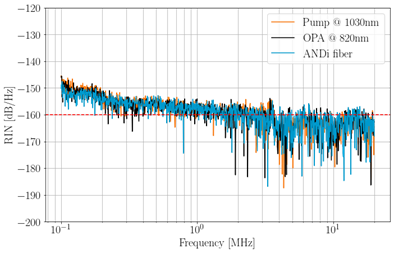}
\caption{\label{fig:psd} Power spectral density (PSD) of the pump laser (1030\,nm), the SC, and the amplified SC (850\,nm). The dashed red line indicates the shot-noise floor.}
\end{figure}

The fiber is a PM-ANDi PCF from NKT Photonics (Item Number NL-1050-NEG-PM). This fiber has a core diameter of 2.4 µm, a relative hole diameter of d/L = 0.45, a small hole-to-hole pitch of L = 1.44 mm, and a nonlinear coefficient of 27\,W$^{-1}$km$^{-1}$. The PM property of the PCF is mainly stress-rod induced and it has a slightly elliptic core, which together gives it a high birefringence of 4.7\,10$^{-4}$ at 1030\,nm. The dispersion has an all-normal dispersion parabolic profile with a peak at -13 ps/nm/km at 1040\,nm. 
The PM-ANDi fiber exhibits a typical throughput of 50\%, and the generated SC spans from 0.75\,µm to 1.3\,µm. Contrary to non-PM ANDi fibers, the polarization state of the SC is strictly linear. After frequency doubling in a 3.7\,mm LBO crystal under type I noncritical phase-matching, the second harmonic of the pump laser is sent to pump the single-pass OPA in a 2.8\,mm type-I LBO crystal cut at $\phi$=90° and $\theta$=11°. Both the SHG pump and SC beams are focused on the OPA crystal with a beam waist of 10\,µm at $1/e^2$. The pump-SC delay is controlled by a motorized delay line. The intrinsic dispersion of the generated SC combined to the dispersion of the collimation/focusing lenses stretch the SC pulse duration to a few ps, allowing to select the amplified bandwidth with the pump-SC delay. A set of dichroic mirrors isolates the amplified signal wavelengths (<1.03\,µm) from the residual 515\,nm and the idler beam (>1.03\,µm). 
When pumped with 9\,W of green light, the average power of the amplified signal reaches ~2\,W from 800\,nm to 950\,nm (\autoref{fig:OPA_spectrum}), which corresponds to quantum conversion efficiency in the range of 35-40\%. The pulse duration at the output of the OPA is about 50-100\,fs FTL, depending on the wavelength.
To quantify the RIN of the pump and amplified SC we follow \cite{audier20} and use an InGaAs photodiode, a 50\,$\Omega$ load, a low-pass filter, a voltage amplifier and a lock-in amplifier (LIA) in sweeper mode. The observable is the power spectral density (PSD) of the voltage at the input of the LIA. The photodiode (DET10N2, Thorlabs) is a free-space, biased (5\,V), DC-coupled photodiode with a rise time of 5\,ns, a bandwidth of 70\,MHz, a saturation photocurrent specified to 5\,mA. A 25\,MHz low-pass filter (BLP-25+, 15542, Mini-circuits) is used to dampen the 40\,MHz repetition rate and harmonics of the optical pulse train. The insertion loss is about 0.5\,dB, with a -3dB cut-off at 28\,MHz and a rejection of 34\,dB at 40\,MHz. The dynamic range of the LIA being not large enough to reach the targeted power spectral density, a voltage amplifier is inserted between the low-pass filter and the LIA. We use an AC-coupled, low-noise amplifier (ZFL-500LN-BNC+, Mini-Circuits) with a 28\,dB gain over 0.1-500\,MHz.
The LIA (HF2LI, Zurich Instruments) has a bandwidth of 50\,MHz (DC-50 MHz), and a sampling rate of 210\,MSa/s. In order to take into account the frequency-dependent gain/loss of detection chain we first measure, independently, the transfer function of the LIA, low-pass filter and amplifier. The PSD at the photodetector level is then deduced from the measured PSD by dividing by the product of the transfer function. The RIN is then assessed by dividing the calculated PSD by the average photocurrent power (assuming a 50\,Ohm load). Classically, the RIN associated to the shot noise of an electric current is : \begin{equation}\label{eq:shot-noise}
    RIN_\text{shot-noise} = \frac{2 q}{I_{avg}}
\end{equation}
where $q$ is the electronic charge and $I_{avg}$ is the average detected photocurrent. For $I_{avg}$=3.7\,mA (experimental value), the RIN of the electrical shot noise is expected to be of -160\,\text{ dB/Hz}. The PSD of the pump beam (1030\,nm), SC (broadband) and amplified SC (820\,nm) is compared in \autoref{fig:psd}. The SC prior to amplification displays a RIN slightly higher than the pump itself. However, the RIN of the amplified SC seems lower, which is compatible with the deep saturation level of the OPA. All three curves reach the shot-noise level (for the aforementioned photocurrent) at frequencies >2\,MHz. 

\section{Proof of concept: SRS imaging}

SRS microscopy is demonstrated by mapping the O-H bond at 2900 cm$^{-1}$ in olive oil droplets immersed in water. The signal output from the OPA (tuned to 790\,nm) is first modulated at 10\,MHz by an acousto-optic modulator (MT200, AA Opto Electronic) and collinearly recombined with part of the 1030\,nm beam by a dichroic mirror. The beams are then directed to 2D-galvoscanner (Saturn 9B, ScannerMAX, 10\,mm aperture) and focused by an f=110\,mm lens to $\simeq$10\,µm FWHM. At focus, the pump power for SRS (790\,nm) is 300\,mW and the SRS Stokes power (1030\,nm) is 100\,mW. The detection chain is made up of a condenser, two high-pass type optical spectral filters (FELH1000, Thorlabs), an InGaAs photodiode (DET10N2, Thorlabs). The photodiode signal is filtered using two low-pass RF filters (CLPFL-0025-BNC from Crystek and BLP-25+ from Mini-circuits), to filter out the pulse structure while retaining rapid variations (<25\,MHz). The resulting RF signal is analyzed by the LIA, configured to demodulate at AOM frequency (10\,MHz). The photodiode operates in a linear regime, just below the saturation threshold. For the selected scan parameters (40x40 pixels in 10\,s), the integration time per pixel is ~62.5 ms. The scanned area is 400\,µm$\times$400\,µm with 40$\times$40 steps. The SRS signal vanishes when the pump-Stokes relative delay exceeds the pulse duration and when the wavelength of the SRS pump is shifted to longer wavelengths. Note that this breadboard experiment was done to demonstrate the ability of the developed laser source to perform SRS and that there is much more to expect in terms of SRS image quality when the laser source will be implemented into a proper microscope. 
\begin{figure}[h]
\centering
\includegraphics[width=0.7\textwidth]{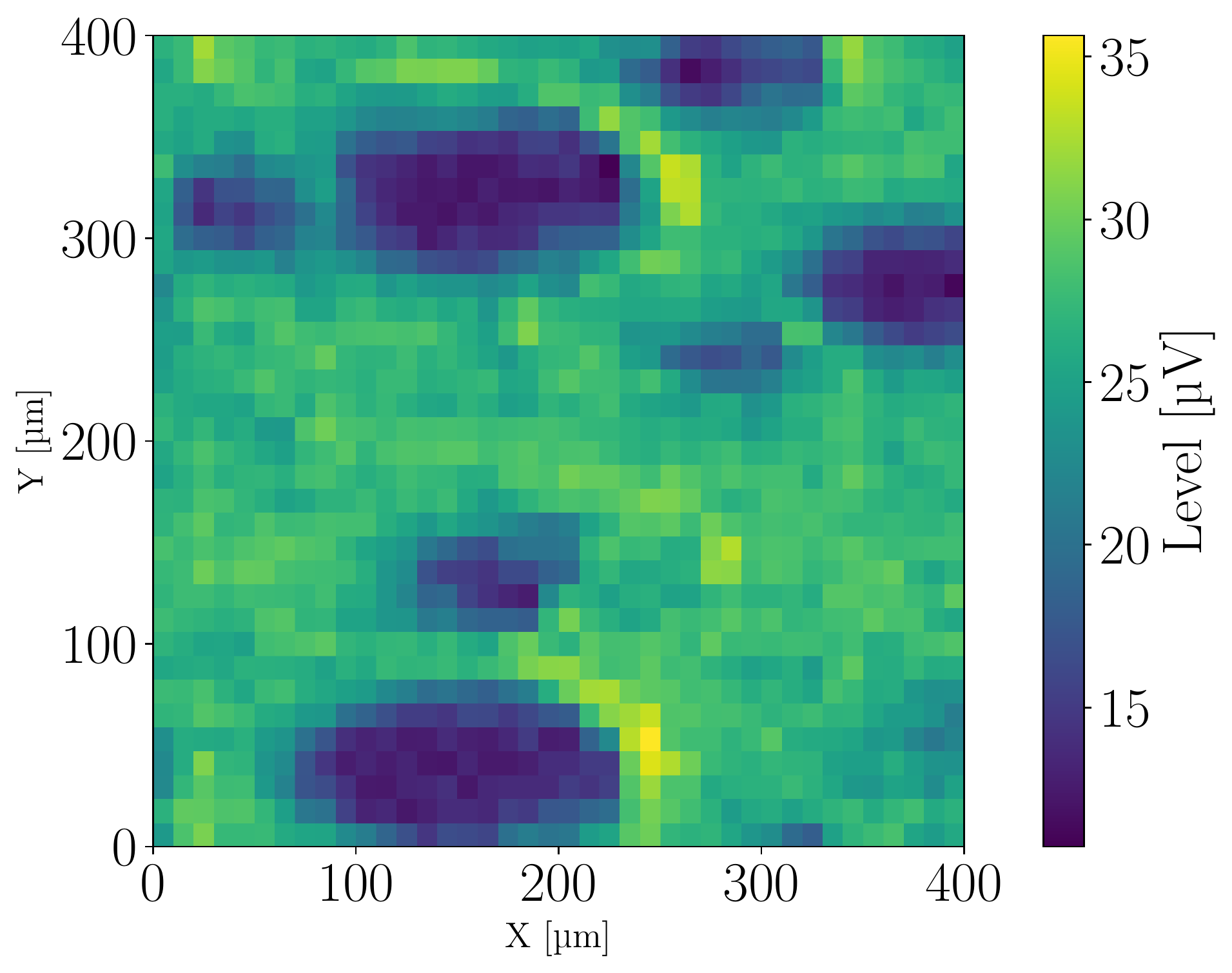}
\caption{\label{fig:srs} SRS image of droplets of vinegar in olive oil at 2900 cm$^{-1}$}.
\end{figure}
\autoref{fig:srs} shows the SRS image at 2900\,cm$^{-1}$. In dark are the vinegar droplets and in light green the SRS signal from olive oil. With the current setup, the signal-to-noise is about 2. This weak value may be explained by the short pulse duration of the pump and Stokes pulses ($\simeq$75\,fs and $\simeq$150\,fs at 790\,nm and 1030\,nm respectively) but also the large beam size on the same. Last, the angular chirp introduced by the AOM was not compensated for in this experiment and the pump beam (790\,nm) was strongly elliptic at the focus. The large spectral bandwidth of the pump and Stokes pulses will be exploited to implement spectral focusing in future work, in order to improve significantly the signal-to-noise of the measurement. 

\section{Conclusion}
In summary, we have demonstrated an efficient single-pass OPA based on a single Kerr-lens-mode-locked Ytterbium laser. The OPA pump is provided by the SHG of the Ytterbium laser at 515\,nm, while the OPA seed is provided by an octave-spanning supercontinuum excited at 1030\,nm. The OPA gain bandwidth nicely overlaps with the SCG spectrum, providing a continuous wavelength tunable range throughout the near-infrared regime of 800-950\,nm for the signal beam. Over 35\% internal conversion efficiency is achieved at 40\,MHz repetition rate in the OPA, which corresponds to an output average power of 1.5-2\,W over the full tunability range. The relative intensity of the OPA matches that of the pump laser and we demonstrate a shot-noise-limited RIN above 2\,MHz. The low noise level, the pulse energy and the high repetition rate of this broadband tunable source are comparable to or exceed the characteristics of  synchronously pumped optical parametric oscillators. As a proof of concept, we image vinegar droplets in oil using stimulated Raman scattering.

\begin{backmatter}
\bmsection{Funding}
This work is supported by European Union through the European Regional Development Fund (ERDF) forming part of the Programme Opérationnel FEDER-FSE Provence Alpes Côte d’Azur 2014-2020.

\bmsection{Acknowledgments}
The authors thank NKT Photonics for the loan the PM-ANDI PCF Fiber Item (Number NL-1050-NEG-PM).

\bmsection{Disclosures}
The authors declare no conflicts of interest.

\bmsection{Data Availability Statement}
Data can be made available upon reasonable request.

\end{backmatter}


\bibliography{biblio}






\end{document}